\documentclass[a4paper]{IEEEtran}
\pdfoutput=1 
\usepackage{color}
\usepackage{leftidx}
\usepackage{cite}
\usepackage{microtype}
\usepackage{subfigure}
\usepackage[font=small]{caption}
\usepackage{graphicx,amssymb,amstext,amsmath}
\usepackage{cleveref}
\usepackage[ruled,vlined,lined,ruled,linesnumbered]{algorithm2e}
\begin{document}

\title{Stable Online Computation Offloading via Lyapunov-guided Deep Reinforcement Learning}

\author{Suzhi~Bi$^*$, Liang~Huang$^\dagger$, Hui Wang$^*$, and Ying-Jun~Angela~Zhang$^\ddag$\\
$^*$College of Electronics and Information Engineering, Shenzhen University, Shenzhen, China\\
$^\dagger$College of Computer Science and Technology, Zhejiang University of Technology, Hangzhou, China\\
$^\ddag$Department of Information Engineering, The Chinese University of Hong Kong, Hong Kong SAR\\
E-mail:~\{bsz,wanghsz\}@szu.edu.cn, liang.huang08@gmail.com, yjzhang@ie.cuhk.edu.hk \vspace{-2ex}}

\maketitle

\vspace{-10pt}

\begin{abstract}
In this paper, we consider a multi-user mobile-edge computing (MEC) network with time-varying wireless channels and stochastic user task data arrivals in sequential time frames. In particular, we aim to design an online computation offloading algorithm to maximize the network data processing capability subject to the long-term data queue stability and average power constraints. The online algorithm is practical in the sense that the decisions for each time frame are made without the assumption of knowing future channel conditions and data arrivals. We formulate the problem as a multi-stage stochastic mixed integer non-linear programming (MINLP) problem that jointly determines the binary offloading (each user computes the task either locally or at the edge server) and system resource allocation decisions in sequential time frames. To address the coupling in the decisions of different time frames, we propose a novel framework, named LyDROO, that combines the advantages of Lyapunov optimization and deep reinforcement learning (DRL). Specifically, LyDROO first applies Lyapunov optimization to decouple the multi-stage stochastic MINLP into deterministic per-frame MINLP subproblems of much smaller size. Then, it integrates model-based optimization and model-free DRL to solve the per-frame MINLP problems with very low computational complexity. Simulation results show that the proposed LyDROO achieves optimal computation performance while satisfying all the long-term constraints. Besides, it induces very low execution latency that is particularly suitable for real-time implementation in fast fading environments.
\end{abstract}

\vspace{-2ex}

\IEEEpeerreviewmaketitle

\section{Introduction}
The emerging mobile-edge computing (MEC) technology allows WDs to offload intensive computation tasks to the edge server (ES) in the vicinity to reduce the computation energy and time cost \cite{2017:Mao}. Compared to the naive scheme that offloads all the tasks for edge execution, \emph{opportunistic computation offloading}, which dynamically assigns tasks to be computed either locally or at the ES, has shown significant performance improvement under time-varying network conditions, such as wireless channel gains\cite{2013:Wu}, harvested energy level\cite{2016:You}, service availability \cite{2020:Bi2}, and task input-output dependency \cite{2020:Yan}. In general, it requires solving a mixed integer non-linear programming (MINLP) that jointly determines the binary offloading (i.e., either offloading the computation or not) and the communication/computation resource allocation (e.g., task offloading time and local/edge CPU frequencies) decisions \cite{2020:Yan,2018:Bi,2017:Dinh}. To tackle the prohibitively high computational complexity of solving the MINLP problems, many works have proposed reduced-complexity sub-optimal algorithms, such as local-search based heuristics \cite{2018:Bi,2020:Yan}, decomposition-oriented search \cite{2018:Bi}, and convex relaxations of the binary variables \cite{2019:Du,2017:Dinh}, etc. However, aside from performance losses, the above algorithms still require a large number of numerical iterations to produce a satisfying solution. It is therefore too costly to implement the conventional optimization algorithms in a highly dynamic MEC environment, where the MINLP needs to be frequently re-solved once the system parameters, such as wireless link quality, vary.

The recent development of data-driven \emph{deep reinforcement learning} (DRL) provides a promising alternative to tackle the online computation offloading problem. In a nutshell, the DRL framework takes a model-free approach that uses deep neural networks (DNNs) to directly learn the optimal mapping from the ``state" (e.g., time-varying system parameters) to the ``action" (e.g., offloading decisions and resource allocation) to maximize the ``reward" (e.g., data processing rate). Example implementations include using deep Q-learning network (DQN) \cite{2019:Min}, double DQN \cite{2019:Chen}, actor-critic DRL \cite{2019:Wei,2019:Huang}. In particular, our previous work \cite{2019:Huang} proposes a hybrid framework, named DROO (Deep Reinforcement learning-based Online Offloading), to combine the advantages of conventional model-based optimization and model-free DRL methods. The integrated learning and optimization approach leads to more robust and faster convergence of the online training process, thanks to the accurate estimation of reward values corresponding to each sampled action.

Apart from optimizing the computation performance, it is equally important to guarantee stable system operation, such as data queue stability and average power consumption. However, most of the existing DRL-based methods do not impose long-term performance constraints (e.g., \cite{2019:Min,2019:Chen,2019:Wei,2019:Huang}). Instead, they resort to heuristic approaches that discourage unfavorable actions in each time frame by introducing penalty terms related to, for example, packet drop events \cite{2019:Chen} and energy consumption \cite{2019:Min}. A well-known framework for online joint utility maximization and stability control is \emph{Lyapunov optimization} \cite{2010:Neely}. It decouples a multi-stage stochastic optimization to sequential per-stage deterministic subproblems, while providing theoretical guarantee to long-term system stability. Some recent works have applied Lyapunov optimization to design computation offloading strategy in MEC networks (e.g., \cite{2019:Du,2016:Mao,2019:Liu1}). However, it still needs to solve a hard MINLP in each per-stage subproblem to obtain the joint binary offloading and resource allocation decisions. To tackle the intractability, some works have designed reduced-complexity heuristics, such as continuous relaxation in \cite{2019:Du} and decoupling heuristic in \cite{2019:Liu1}. This, however, suffers from the similar performance-complexity tradeoff dilemma as in \cite{2020:Yan,2018:Bi,2017:Dinh}.

In this paper, we consider a multi-user MEC network with an ES assisting the computation of $N$ WDs, where the computation task data arrive at the WDs' data queues stochastically in sequential time frames. We aim to design an online computation offloading algorithm, in the sense that the decisions for each time frame are made without knowing the future channel conditions and task arrivals. The objective is to maximize the network data processing capability under the long-term data queue stability and average power constraints. To tackle the problem, we propose a Lyapunov-guided Deep Reinforcement learning (DRL)-based Online Offloading (LyDROO) framework that combines the advantages of Lyapunov optimization and DRL. In particular, we first apply Lyapunov optimization to decouple the multi-stage stochastic MINLP into per-frame deterministic MINLP problems. Then in each frame, we integrate model-based optimization and model-free DRL to solve the per-frame MINLP problems with very low computational complexity. Simulation results show that the proposed LyDROO algorithm converges very fast to the optimal computation rate while meeting all the long-term stability constraints. Compared to a myopic benchmark algorithm that greedily maximizes the computation rate in each time frame, the proposed LyDROO achieves a much larger \emph{stable capacity region} that can stabilize the data queues under much heavier task data arrivals.

\section{System Model and Problem Formulation}
We consider an MEC network with an ES assisting the computation of $N$ WDs in sequential time frames of equal duration $T$. Within the $t$th time frame, we denote $A_{i}^t$ (in bits) as the raw task data arrival at the data queue of the $i$th WD. We assume that $A_{i}^t$ follows an i.i.d. exponential distribution with mean $\lambda_i$, for $i=1,\cdots,N$. We denote the channel gain between the $i$th WD and the ES as $h^t_i$. Under the block fading assumption, $h^t_i$ remains constant within a time frame but varies independently across different frames.

In the $t$th time frame, suppose that a tagged WD $i$ processes $D_{i}^t$ bits data and produces a computation output at the end of the time frame. Within each time frame, we assume that the WDs adopt a binary computation offloading rule \cite{2017:Mao} that process the raw data either locally at the WD or remotely at the ES. The offloading WDs share a common bandwidth $W$ for transmitting the task data to the ES in a TDMA manner. We use a binary variable $x^t_i$ to denote the offloading decision.

When the $i$th WD processes the data locally ($x^t_i=0$), we denote the local CPU frequency as $f^t_i$, which is upper bounded by $f_i^{max}$. The raw data (in bits) processed locally and the consumed energy within the time frame are \cite{2017:Mao}
\begin{equation}
\label{51}
\small
D^{t}_{i,L} = f^t_i T/\phi,\ E^{t}_{i,L} = \kappa \left(f^t_i\right)^3 T, \ \forall x_{i}^t =0,
\end{equation}
respectively. Here, parameter $\phi>0$ denotes the number of computation cycles needed to process one bit of raw data and $\kappa>0$ denotes the computing energy efficiency parameter.

Otherwise, when the data is offloaded for edge execution ($x^t_i=1$), we denote $P_i^t$ as the transmit power constrained by the maximum power $P_i^t \leq P^{max}_i$ and $\tau^t_i T$ as the amount of time allocated to the $i$th WD for computation offloading. Here, $\tau^t_i \in [0,1]$ and $\sum_{i=1}^N \tau^t_i \leq 1$. The energy consumed on data offloading is $E^{t}_{i,O} = P_i^t \tau^t_i T$, such that $P_i^t = E_{i,O}^t/\tau^t_i T$. Similar to \cite{2016:You} and \cite{2018:Bi}, we neglect the delay on edge computing and result downloading such that the amount of data processed at the edge within the time frame is
\begin{equation}
\label{52}
\small
\begin{aligned}
D^{t}_{i,O} = \frac{W\tau^t_i T}{v_u}\log_2\left(1+\frac{E_{i,O}^t h_i^t}{\tau^t_i T N_0}\right),\ \forall x_{i}^t =1,
\end{aligned}
\end{equation}
where $v_u\geq 1$ represents the rate loss due to communication overhead and $N_0$ denotes the noise power.

Let $D^{t}_{i} \triangleq (1 - x^t_i) D^{t}_{i,L} + x^t_i D^{t}_{i,O}$ and $E^t_i \triangleq (1 - x^t_i) E^{t}_{i,L} + x^t_i E^{t}_{i,O}$ denote the bits computed and energy consumed in time frame $t$. We define \emph{computation rate} $r_i^t$ and power consumption $e_i^t$ in the $t$th time frame as
\begin{equation}
\label{14}
\small
\begin{aligned}
r_i^t = \frac{D^{t}_{i}}{T} &=\frac{(1 - x^t_i)f^t_i}{\phi} + x^t_i \frac{W\tau^t_i}{v_u}\log_2\left(1+\frac{e^t_{i,O} h_i^t}{\tau^t_i N_0}\right), \\
e_i^t = \frac{E^{t}_{i}}{T} & = (1 - x^t_i) \kappa \left(f^t_i\right)^3 + x^t_i e^t_{i,O},
\end{aligned}
\end{equation}
where $e^t_{i,O} \triangleq  E^{t}_{i,O}/T $. For simplicity of exposition, we assume $T=1$ without loss of generality in the following derivations.

Let $Q_{i}(t)$ denote the queue length of the $i$th WD at the beginning of the $t$th time frame such that
\begin{equation}
Q_{i}(t+1) = \max\left\{Q_{i}(t) - \tilde{D}^{t}_{i} + A_{i}^t ,0\right\}, \ i=1,2,\cdots,
\end{equation}
where $\tilde{D}^{t}_{i} = \min \left(Q_{i}(t),D^{t}_{i}\right)$ and $Q_{i}(1)=0$. In the following derivation, we enforce the data causality constraint $D^{t}_{i} \leq Q_{i}(t)$, implying that $Q_i(t)\geq 0$ holds for any $t$. Thus, the queue dynamics is simplified as
\begin{equation}
\label{111}
Q_i(t+1) = Q_{i}(t) - D^{t}_{i} + A_{i}^t,\ \ i=1,2,\cdots.
\end{equation}
If $\lim_{K\rightarrow \infty} \frac{1}{K} \mathsmaller\sum_{t=1}^K \mathbb{E} \left[Q_i(t)\right] <\infty$ holds for $Q_{i}(t)$, we refer to the queue as \emph{strongly stable}, where the expectation is taken with respect to the random channel fading and task data arrivals \cite{2010:Neely}. By the Little's law, a strongly stable data queue translates to a finite processing delay of each task data bit.

In this paper, we aim to design an online algorithm to maximize the long-term average weighted sum computation rate of all the WDs under the data queue stability and average power constraints. We denote $\mathbf{x}^t = \left[x_1^t,\cdots,x_N^t\right]$, $\boldsymbol{\tau}^t = \left[\tau_1^t,\cdots,\tau_N^t\right]$, $\mathbf{f}^t = \left[f_1^t,\cdots,f_N^t\right]$ and $\mathbf{e}_O^t = \left[e_{1,O}^t,\cdots,e_{N,O}^t\right]$, and let $\mathbf{x} = \left\{\mathbf{x}^t\right\}_{t=1}^K$, $\boldsymbol{\tau} = \left\{\boldsymbol{\tau}^t\right\}_{t=1}^K$, $\mathbf{f}= \left\{\mathbf{f}^t\right\}_{t=1}^K$ and $\mathbf{e}_O = \left\{\mathbf{e}_O^t \right\}_{t=1}^K$. We formulate the problem as a multi-stage stochastic MINLP

\begin{subequations}
   \label{6}
   \small
   \begin{align}
    & \underset{\mathbf{x}, \boldsymbol{\tau},\mathbf{f},\mathbf{e}_O}{\text{maximize}} \ \  \lim_{K\rightarrow \infty} 1/K \cdot \mathsmaller\sum_{t=1}^K \mathsmaller\sum_{i=1}^N c_i r_i^t \label{61}\\
    & \text{subject to} &  & \\
    & \mathsmaller\sum_{i=1}^N \tau^t_i  \leq 1, \ \forall t, \ \ x^t_i \in \left\{0,1\right\}, \forall i,t, \label{62}\\
    &  \frac{(1-x^t_i)f^t_i}{\phi}  + \frac{x^t_iW\tau^t_i}{v_u}\log_2\left(1+\frac{e^t_{i,O} h_i^t}{\tau^t_i N_0}\right) \leq Q_i(t) , \ \forall t, i, \label{65}\\
    &  \lim_{K\rightarrow \infty} 1/K \cdot \mathsmaller\sum_{t=1}^K \mathbb{E}\left[(1 - x^t_i) \kappa \left(f^t_i\right)^3 + x^t_i e^t_{i,O}\right] \leq \gamma_i,\  \forall i,\label{63}\\
    &  \lim_{K\rightarrow \infty} 1/K \cdot \mathsmaller\sum_{t=1}^K \mathbb{E} \left[Q_i(t)\right] <\infty, \forall i, \label{64}\\
    & \tau^t_i, f^t_i,e^t_{i,O} \geq 0, \ f^t_i \leq f^{max}_i,\ e^t_{i,O}\leq P^{max}_i \tau^t_i, \forall i,t.
   \end{align}
\end{subequations}
Here, $c_i$ denotes the fixed weight of the $i$th WD. (\ref{62}) denotes the offloading time constraint. Notice that $\tau^t_i = e^t_{i,O}=0$ must hold at the optimum if $x^t_i=0$. Similarly, $f^t_i=0$ must hold if $x^t_i=1$. (\ref{65}) corresponds to the data causality constraint. (\ref{63}) corresponds to the average power constraint and $\gamma_i>0$ is the power threshold. (\ref{64}) are the data queue stability constraints. Under the stochastic channels and data arrivals, it is hard to satisfy the long-term constraints when the decisions are made in each time frame without future knowledge. Besides, the fast-varying channel condition requires real-time decision-making in each short time frame, e.g., within the channel coherence time. In the following, we propose a novel LyDROO framework that solves (\ref{6}) with both high robustness and efficiency.

\section{Lyapunov-based Multi-Stage Decoupling}
In this section, we apply the Lyapunov optimization to decouple (\ref{6}) into per-frame deterministic problems. To cope with the average power constraints (\ref{63}), we introduce $N$ virtual energy queues $\left\{Y_i(t)\right\}_{i=1}^N$, one for each WD. Specifically, we set $Y_i(1) =0$ and update the queue as
\begin{equation}
\label{112}
Y_i(t+1) = \max \left(Y_i(t) + \nu e^t_i - \nu \gamma_i, 0 \right),
\end{equation}
for $i=1,\cdots,N,\ t=1,2,\cdots$, where $e^t_i$ in (\ref{14}) is the energy consumption at the $t$th time frame and $\nu$ is a positive scaling factor. Intuitively, when the virtual energy queues are stable, the average power consumption $e_i^t$ does not exceed $\gamma_i$, and thus the constraints in (\ref{63}) are satisfied.

We define $\mathbf{Z}(t) = \left\{\mathbf{Q}(t),\mathbf{Y}(t)\right\}$ as the total queue backlog, where $\mathbf{Q}(t) = \left\{Q_i(t)\right\}_{i=1}^N$ and $\mathbf{Y}(t) = \left\{Y_i(t)\right\}_{i=1}^N$. Then, we introduce the Lyapunov function $L\left( \mathbf{Z}(t)\right)$ and Lyapunov drift $\Delta L\left( \mathbf{Z}(t)\right)$ as  \cite{2010:Neely}
\begin{equation}
\small
\begin{aligned}
L\left( \mathbf{Z}(t)\right) &= 0.5\left( \mathsmaller\sum_{i=1}^N Q_i(t)^2 +  \mathsmaller\sum_{i=1}^N Y_i(t)^2\right), \\ \Delta L\left( \mathbf{Z}(t)\right) &= \mathbb{E}\left\{L\left(\mathbf{Z}(t+1)\right)-L\left(\mathbf{Z}(t)\right)| \mathbf{Z}(t) \right\}.
\end{aligned}
\end{equation}
To maximize the time average computation rate while stabilizing $\mathbf{Z}(t)$, we use the drift-plus-penalty minimization approach \cite{2010:Neely}. Specifically, we seek to minimize an upper bound on the following drift-plus-penalty expression at every time frame $t$:
\begin{equation}
\label{57}
\small
\Lambda\left(\mathbf{Z}(t)\right) \triangleq \Delta L\left( \mathbf{Z}(t)\right) - V \cdot \mathsmaller\sum_{i=1}^N \mathbb{E} \left\{ c_i r_i^t|\mathbf{Z}(t)\right\},
\end{equation}
where $V >0$ is an ``importance" weight to scale the penalty. The following Theorem $1$ derives an upper bound of $\Lambda\left(\mathbf{Z}(t)\right)$.

\textbf{Theorem 1:} Given any queue backlog $\mathbf{Z}(t)$, the drift-plus-penalty in (\ref{57}) is upper bounded as
\begin{equation}
\label{12}
\small
\begin{aligned}
&\Lambda\left(\mathbf{Z}(t)\right) \leq \hat{B}  + \mathsmaller\sum_{i=1}^N   Q_{i}(t) \mathbb{E}\left[\left(A_{i}^t-D^{t}_{i}\right)| \mathbf{Z}(t)\right] \\
&+  \mathsmaller\sum_{i=1}^N \left\{Y_i(t) \mathbb{E}\left[e^t_i -\gamma_i| \mathbf{Z}(t)\right] - V \mathbb{E} \left[ c_i r_i^t|\mathbf{Z}(t)\right] \right\},
\end{aligned}
\end{equation}
where $\hat{B}$ is a finite constant.

\emph{Proof}: Due to the page limit, please refer the detailed proof in the online technical report \cite{2020:Bi1}. $\hfill \blacksquare$

In the $t$th time frame, we apply the technique of \emph{opportunistic expectation minimization}\cite{2010:Neely}. That is, we observe the queue backlogs $\mathbf{Z}(t)$ and decide the joint offloading and resource allocation control action accordingly to minimize the upper bound in (\ref{12}). By removing the constant terms from the observation at the beginning of the $t$th time frame, the algorithm decides the actions by maximizing the following:
\begin{equation}
\mathsmaller\sum_{i=1}^N \left(Q_{i}(t) + V c_i\right) r^{t}_{i} - \mathsmaller\sum_{i=1}^N Y_i(t)e^t_i,
\end{equation}
where $r^{t}_{i}$ and $e^t_i$ are in (\ref{14}). Intuitively, it tends to increase the computation rates of WDs that have a long data queue backlog or a large weight, while penalizing those that have exceeded the average power threshold. We introduce an auxiliary variable $r_{i,O}^t$ for each WD $i$ and denote $\mathbf{r}_{O}^t = \left\{r_{i,O}^t\right\}_{i=1}^N$. Taking into account the per-frame constraints, we solve the following deterministic per-frame subproblem in the $t$th time frame
\begin{subequations}
   \label{7}
   \small
   \begin{align}
    & \underset{\mathbf{x}^t, \boldsymbol{\tau}^t,\mathbf{f}^t,\mathbf{e}_O^t, \mathbf{r}_{O}^t}{\text{maximize}} \    \mathsmaller\sum_{i=1}^N \left(Q_{i}(t) + V c_i\right) r^{t}_{i} - \mathsmaller\sum_{i=1}^N Y_i(t)e^t_i \label{80}\\
    & \text{subject to} \\
    & \qquad \mathsmaller \sum_{i=1}^N \tau^t_i  \leq 1, \label{25}\\
    & \qquad f^t_i /\phi \leq  Q_i(t), \ r_{i,O}^t \leq  Q_i(t),\ \forall i, \label{78}\\
    & \qquad r_{i,O}^t \leq \frac{ W\tau^t_i}{v_u}\log_2\left(1+\frac{e^t_{i,O} h_i^t}{\tau^t_i N_0}\right), \ \forall i, \label{79}\\
    & \qquad f^t_i \leq f^{max}_i,\ e^t_{i,O}\leq P^{max}_i \tau^t_i, \ \forall i,\\
    & \qquad x^t_i \in \left\{0,1\right\},\ \tau^t_i, f^t_i,e^t_{i,O} \geq 0, \ \forall i. \label{85}
   \end{align}
\end{subequations}
Notice that the above constraints (\ref{78}) and (\ref{79}) are equivalent to (\ref{65}), because there is exactly one non-zero term in the left-hand side of (\ref{65}) at the optimum. It can be shown that, for a feasible (\ref{6}) under the data arrival rates and power constraints, we can satisfy all long-term constraints in (\ref{6}) by solving the per-frame subproblems in an online fashion, where the detailed proof is omitted due to the page limit. Then, the remaining difficulty lies in solving the MINLP (\ref{7}) in each time frame.

\begin{figure*}
\centering
\includegraphics[width=0.7 \textwidth]{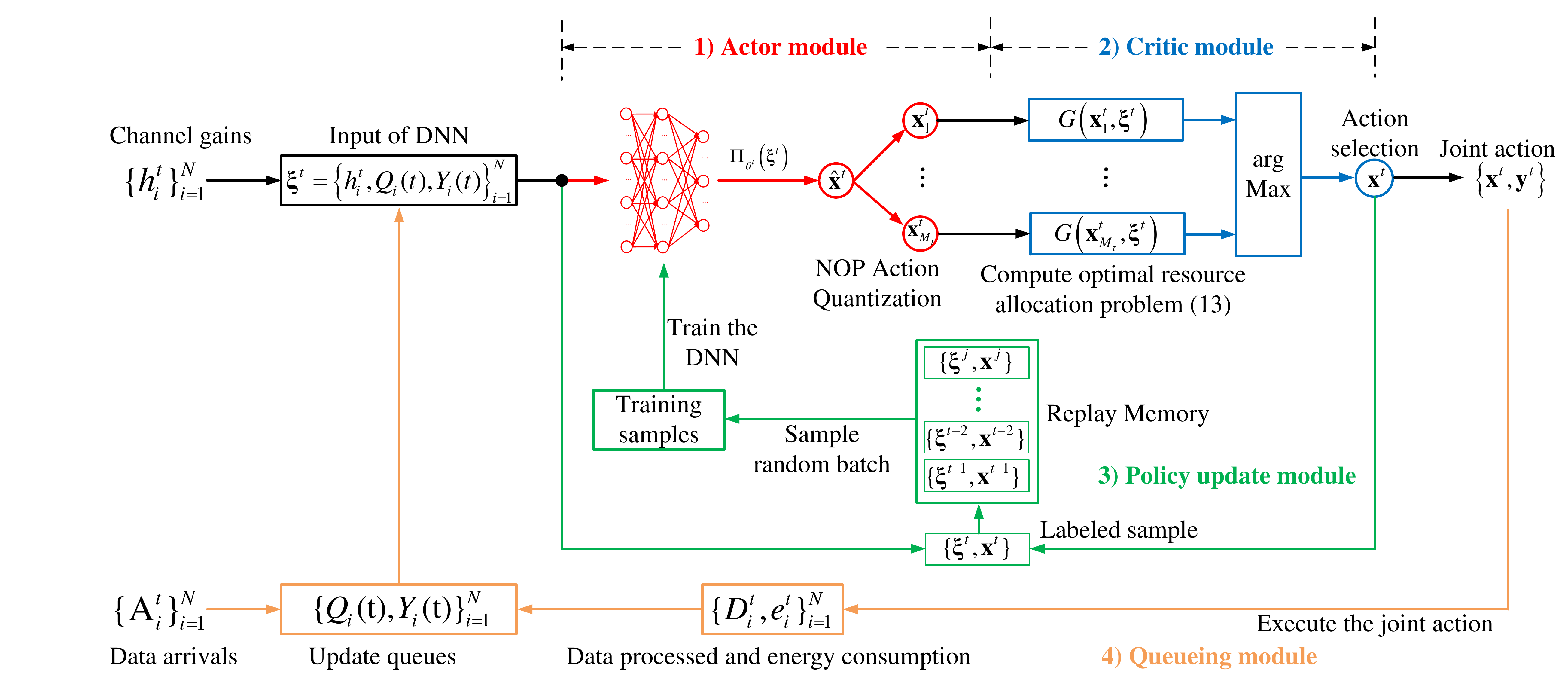}
\caption{The schematics of the proposed LyDROO algorithm.}
\label{fig:network}
\end{figure*}

\section{Lyapunov-guided DRL for Online Offloading}
Recall that to solve (\ref{7}) in the $t$th time frame, we observe $\boldsymbol{\xi}^t \triangleq \left\{h^t_i, Q_i(t), Y_i(t)\right\}_{i=1}^N$, consisting of the channel gains $\left\{h^t_i\right\}_{i=1}^N$ and the system queue states $\left\{Q_i(t), Y_i(t)\right\}_{i=1}^N$, and accordingly decide the control action $\left\{\mathbf{x}^t,\mathbf{y}^t\right\}$, including the binary offloading decision $\mathbf{x}^t$ and the continuous resource allocation $\mathbf{y}^t \triangleq \left\{\tau_i^t, f_i^t, e_{i,O}^t, r_{i,O}^t\right\}_{i=1}^N$.

Suppose that the value of $\mathbf{x}^t$ in (\ref{7}) are given, we denote the index set of users with $x^t_i=1$ as $\mathcal{M}_1^t$, and the complementary user set with $x^t_i=0$ as $\mathcal{M}_0^t$. Then, the remaining optimal resource allocation problem to optimize $\mathbf{y}^t$ is as following
\begin{subequations}
\allowdisplaybreaks
   \label{8}
   \small
   \begin{align}
    & \underset{\boldsymbol{\tau}^t,\mathbf{f}^t,\mathbf{e}_O^t, \mathbf{r}_{O}^t}{\text{maximize}}\   \mathsmaller \sum_{j\in\mathcal{M}_0^t} \left\{a_j^t f_j^t/\phi -  Y_j(t) \kappa \left(f_j^t\right)^3 \right\} \\
    & \qquad  \qquad + \mathsmaller \sum_{i\in\mathcal{M}_1^t} \left\{a_i^t r_{i,O}^t - Y_i(t) e_{i,O}^t \right\} \\
    & \text{subject to} \\
    & \qquad \mathsmaller \sum_{i\in \mathcal{M}_1} \tau_i^t  \leq 1, \ e_{i,O}^t\leq P^{max}_i \tau_i^t, \forall i \in \mathcal{M}_1^t,\\
    & \qquad f_j^t /\phi \leq  Q_j(t), f_j^t \leq f^{max}_j, \ \forall  j \in \mathcal{M}_0^t, \\
    & \qquad  r_{i,O}^t \leq  Q_i(t), \ \forall i \in  \mathcal{M}_1^t \\
    & \qquad r_{i,O}^t \leq \frac{ W\tau_i^t}{v_u}\log_2\left(1+\frac{e_{i,O}^t h_i^t}{\tau_i^t N_0}\right), \ \forall i \in  \mathcal{M}_1^t, \label{73}
    \end{align}
\end{subequations}
where $a_i^t \triangleq Q_{i}(t) + V c_i$ is a parameter. A close observation shows that although (\ref{7}) is a non-convex optimization problem, the resource allocation problem (\ref{8}) is in fact a convex problem if $\mathbf{x}^t$ is fixed. Here, we denote $G\left(\mathbf{x}^t,\boldsymbol{\boldsymbol{\xi}}^t\right)$ as the optimal value of (\ref{7}) by optimizing $\mathbf{y}^t$ given the offloading decision $\mathbf{x}^t$ and parameter $\boldsymbol{\boldsymbol{\xi}}^t$, i.e., the optimal value of (\ref{8}), which can be efficiently obtained from off-the-shelf or customized convex algorithms. In this sense, solving (\ref{7}) is equivalent to finding the optimal offloading decision $\left(\mathbf{x}^t\right)^*$, where
\begin{equation}
\label{94}
    \left(\mathbf{x}^t\right)^* =  \arg \underset{\mathbf{x}^t \in \{0,1\}^N}{\text{maximize}} \ \ \ G\left(\mathbf{x}^t,\boldsymbol{\boldsymbol{\xi}}^t\right).
\end{equation}
In general, obtaining $\left(\mathbf{x}^t\right)^*$ requires enumerating $2^N$ offloading decisions, which leads to significantly high computational complexity even when $N$ is moderate (e.g., $N=10$). Other search based methods, such as branch-and-bound and block coordinate descent, are also time-consuming when $N$ is large. In practice, neither method is applicable to online decision-making under fast-varying channel condition. Leveraging the DRL technique, we propose a LyDROO algorithm to construct a policy $\pi$ that maps from the input $\boldsymbol{\boldsymbol{\xi}}^t$ to the optimal action $\left(\mathbf{x}^t\right)^*$, i.e., $\pi : \boldsymbol{\boldsymbol{\xi}}^t \mapsto \left(\mathbf{x}^t\right)^*$, with very low complexity, e.g., tens of milliseconds execution delay for $N=10$. As illustrated in Fig.~\ref{fig:network}, LyDROO consists of four main modules that operate in a sequential and iterative manner as detailed below.

\subsubsection{Actor Module} The actor module consists of a DNN and an action quantizer. Here, we apply a fully-connected multi-layer perceptron of two hidden layers with $120$ and $80$ hidden neurons, respectively. Besides, we use a sigmoid activation function at the output layer. At the beginning of the $t$th time frame, we denote the parameter of the DNN as $\boldsymbol{\theta}^t$, which is randomly initialized following the standard normal distribution when $t=1$. With the input $\boldsymbol{\boldsymbol{\xi}}^t$, the DNN outputs a relaxed offloading decision $\mathbf{\hat{x}}^t \in [0,1]^{N}$, where the input-out relation is expressed as
\begin{equation}
\label{DNN}
\Pi_{\boldsymbol{\theta}^t}: \boldsymbol{\boldsymbol{\xi}}^t \mapsto \mathbf{\hat{x}}^t = \left\{\hat{x}^t_i \in [0,1], i=1,\cdots,N \right\}.
\end{equation}
We then quantize the continuous $\mathbf{\hat{x}}^t$ into $M_t$ feasible candidate binary offloading actions, denoted as
\begin{equation}
\label{quantize}
\Upsilon_{M_t}: \mathbf{\hat{x}}^t \mapsto  \Omega^t = \left\{\mathbf{x}^t_j | \mathbf{x}^t_j \in \{0,1\}^{N}, j=1,\cdots,M_t \right\},
\end{equation}
where $\Omega_t$ denotes the set of candidate offloading actions in the $t$th time frames. Notice that the number of binary actions $M_t = |\Omega_t|$ is a time-dependent design parameter.

A good quantization function should balance the \emph{exploration-exploitation tradeoff} in generating the offloading action to ensure good training convergence. Intuitively, $\left\{\mathbf{x}^t_j\right\}$'s should be close to $\mathbf{\hat{x}}^t$ (measured by Euclidean distance) to make effective use of the DNN's output and meanwhile sufficiently separate to avoid premature convergence to sub-optimal solution in the training process. Here, we apply a noisy order-preserving (NOP) quantization method, which can generate any $M_t\leq 2N$ candidate actions. The NOP method generates the first $M_t/2$ actions ($M_t$ is assumed an even number) by applying the order-preserving quantizer (OPQ) in \cite{2019:Huang} to $\hat{\mathbf{x}}^t$.\footnote{Please refer to \cite{2020:Bi1} for detailed quantization procedures.} To obtain the remaining $M_t/2$ actions, we first generate a noisy version of $\hat{\mathbf{x}}^t$ denoted as $\tilde{\mathbf{x}}^t = \text{Sigmoid}\left(\hat{\mathbf{x}}^t + \mathbf{n}\right)$, where the random Gaussian noise $\mathbf{n}\sim \mathcal{N}\left(\mathbf{0},\mathbf{I}_{N}\right)$ with $\mathbf{I}_{N}$ being an identity matrix, and $\text{Sigmoid}\left(\cdot\right)$ is the element-wise Sigmoid function that bounds each entry of $\tilde{\mathbf{x}}^t$ within $(0,1)$. Then, we produce the remaining $M_t/2$ actions $\mathbf{x}^t_m$, for $m= M_t/2+1 ,\cdots, M_t$, by applying the OPQ to $\tilde{\mathbf{x}}^t$.

\subsubsection{Critic Module} Followed by the actor module, the critic module evaluates $\left\{\mathbf{x}^t_i\right\}$ and selects the best offloading action $\mathbf{x}^t$. LyDROO leverages the model information to evaluate the binary offloading action by analytically solving the optimal resource allocation problem, which leads to more robust and faster convergence of the DRL training process. Specifically, LyDROO selects the best action $\mathbf{x}^t$ as
\begin{equation}
\label{109}
\mathbf{x}^t = \arg \max_{\mathbf{x}^t_j \in \Omega_t} G\left(\mathbf{x}_j^t,\boldsymbol{\boldsymbol{\xi}}^t\right),
\end{equation}
where $G\left(\mathbf{x}_j^t,\boldsymbol{\boldsymbol{\xi}}^t\right)$ is obtained by optimizing the resource allocation given $\mathbf{x}_j^t$ in (\ref{8}). Intuitively, a larger $M_t = |\Omega_t|$ results in better solution performance, but a higher execution delay. To balance the performance-complexity tradeoff, we propose here an adaptive procedure to set a time-varying $M_t$.

Denote $m_t \in [0,M_t-1]$ as the index of the best action $\mathbf{x}^t \in \Omega_t$. We define $m_t^* = \bmod(m_t, M_t/2)$, which represents the order of $\mathbf{x}^t$ among either the $M_t/2$ noise-free or the noise-added candidate actions. The key idea is that $m_t^*$ gradually decreases as the actor DNN gradually approaches the optimal policy over time. In practice, we set a maximum $M_1 =2N$ initially and update $M_t$ every $\delta_M \geq 1$ time frames as
\begin{equation}
\label{108}
M_{t} = 2\cdot \min\left(\max \left(m_{t - 1}^*,\cdots, m_{t - \delta_M}^* \right)+1,N\right).
\end{equation}
The additional $1$ in the first term within the min operator allows $M_{t}$ to increase over time.

\subsubsection{Policy Update Module} LyDROO uses $\left(\boldsymbol{\boldsymbol{\xi}}^t,\mathbf{x}^t\right)$ as a labeled input-output sample for updating the policy of the DNN. In particular, we maintain a replay memory that only stores the most recent $q$ data samples. In practice, with an initially empty memory, we start training the DNN after collecting more than $q/2$ data samples. Then, the DNN is trained periodically once every $\delta_T$ time slots to avoid model over-fitting. When $\bmod\left(t,\delta_T\right) = 0$, we randomly select a batch of data samples $\left\{\left(\boldsymbol{\boldsymbol{\xi}}^\tau,\mathbf{x}^\tau\right), \tau \in \mathcal{S}^t\right\}$, where $\mathcal{S}^t$ denotes the time indices of the selected samples. We then update the parameter of the DNN by minimizing its average cross-entropy loss function over the data samples. When the training completes, we update the parameter of the actor module in the next time frame to $\boldsymbol{\theta}^{t+1}$.

\subsubsection{Queueing Module}
As a by-product of the critic module, we obtain the optimal resource allocation $\mathbf{y}^t$ associated with $\mathbf{x}^t$. Accordingly, the system executes the joint computation offloading and resource allocation action $\left\{\mathbf{x}^t,\mathbf{y}^t\right\}$, which processes data $\{D_i^t\}_{i=1}^N$ and consumes energy $\{e_i^t\}_{i=1}^N$ as given in (\ref{14}). Based on $\{D_i^t,e_i^t\}_{i=1}^N$ and the data arrivals $\{A_i^t\}_{i=1}^N$ observed in the $t$th time frame, the queueing module then updates the data and energy queues $\left\{Q_i(t+1), Y_i(t+1)\right\}_{i=1}^N$ using (\ref{111}) and (\ref{112}) at the beginning of the $(t+1)$th time frame. With the wireless channel gains observation $\{h_i^{t+1}\}_{i=1}^N$, the system feeds the input parameter $\boldsymbol{\boldsymbol{\xi}}^{t+1} = \left\{h_i^{t+1},Q_i(t+1), Y_i(t+1)\right\}_{i=1}^N$ to the DNN and starts a new iteration from Step 1).

With the above actor-critic-update loop, the DNN consistently learns from the best and most recent state-action pairs, leading to a better policy $\pi_{\boldsymbol{\theta}^t}$ that gradually approximates the optimal mapping to solve (\ref{94}). We summarize the pseudo-code of LyDROO in Algorithm $1$. We can efficiently obtain the best offloading solution $\mathbf{x}^{t}$ in line $8$ by solving convex resource allocation problem in (\ref{8}). Besides, the training process of the DNN in line $10$-$13$ is performed in parallel with the execution of task offloading and computation, and thus does not incur additional training delay overhead.

\begin{algorithm}
\scriptsize
 \SetAlgoLined
 \SetKwData{Left}{left}\SetKwData{This}{this}\SetKwData{Up}{up}
 \SetKwRepeat{doWhile}{do}{while}
 \SetKwFunction{Union}{Union}\SetKwFunction{FindCompress}{FindCompress}
 \SetKwInOut{Input}{input}\SetKwInOut{Output}{output}
 \Input{Parameters $V$, $\left\{\gamma_i, w_i\right\}_{i=1}^N$, $K$, training interval $\delta_T$, $M_t$ update interval $\delta_M$\;}
 \Output{Control actions $\left\{\mathbf{x}^t, \mathbf{y}^t\right\}_{t=1}^K$;}
 Initialize the DNN with random parameters $\boldsymbol{\theta}^{1}$ and empty replay memory, $M_1 \leftarrow 2N$\;
 Initial data and energy queue $Q_i(1) = Y_i(1)=0$, for $i=1,\cdots,N$\;
 \For{$t=1,2,\dots,K$}{
 Observe the input $\boldsymbol{\boldsymbol{\xi}}^t = \left\{h^t, Q_i(t), Y_i(t)\right\}_{i=1}^N$ and update $M_t$ using (\ref{108}) if $\bmod\left(t,\delta_M\right) = 0$\;
 Generate a relaxed offloading action $\hat{\mathbf{x}}^t = \Pi_{\boldsymbol{\theta}^t}\left(\boldsymbol{\boldsymbol{\xi}}^t\right)$ with the DNN\;
 Quantize $\hat{\mathbf{x}}_t$ into binary actions $\left\{\mathbf{x}^t_i| i = 1,\cdots,M_t\right\}$ using the NOP method\;
 Compute $G\left(\mathbf{x}^t_i, \boldsymbol{\boldsymbol{\xi}}^t\right)$ by optimizing $\mathbf{y}^t_i$ in (\ref{8}) for each $\mathbf{x}^t_i$\; \label{step:compute_Q}
 Select the best solution $\mathbf{x}^{t}$ with (\ref{109}) and execute the joint action $\left(\mathbf{x}^{t},\mathbf{y}^t\right)$\;
  Update the replay memory by adding $(\boldsymbol{\boldsymbol{\xi}}^t,\mathbf{x}^{t})$\;
 \If{$\bmod\left(t,\delta_T\right) = 0$}{
   Uniformly sample a data set $\{( \boldsymbol{\boldsymbol{\xi}}^{\tau}, \mathbf{x}^{\tau})\mid \tau \in \mathcal{S}_t\}$ from the memory\;
   Train the DNN with $\{( \boldsymbol{\boldsymbol{\xi}}^{\tau}, \mathbf{x}^{\tau})\mid \tau \in \mathcal{S}_t\}$ and update $\boldsymbol{\theta}^{t}$\;
 }
 $t\leftarrow t+1$\;
 Update $\left\{Q_i(t), Y_i(t)\right\}_{i=1}^N$ based on $\left(\mathbf{x}^{t-1},\mathbf{y}^{t-1}\right)$ and data arrival observation $\left\{A^{t-1}_i\right\}_{i=1}^N$ using (\ref{111}) and (\ref{112}).
 }
 \caption{LyDROO algorithm for solving (\ref{6}).}
\end{algorithm}

\vspace*{-.4cm}

\begin{figure*}
  \centering
  \subfigure{\includegraphics[width=0.23\textwidth]{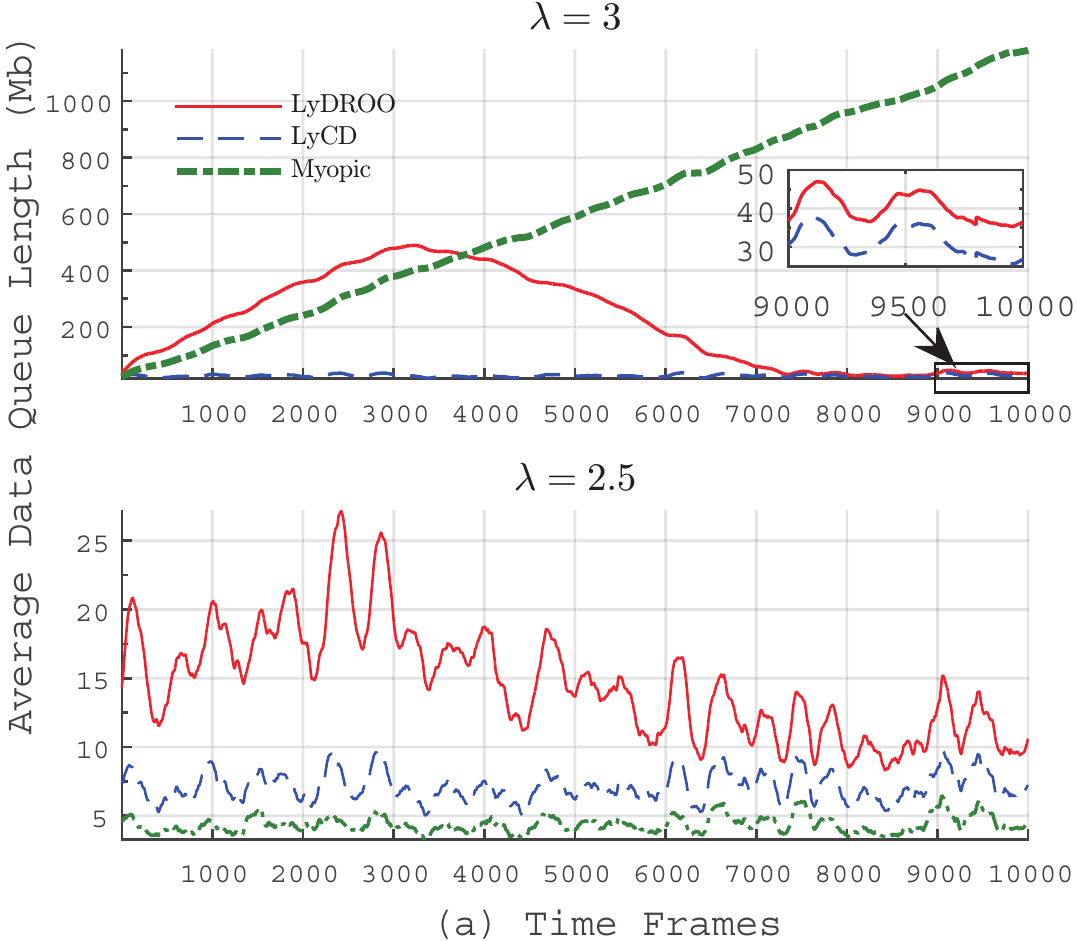}}\quad
  \subfigure{\includegraphics[width=0.23\textwidth]{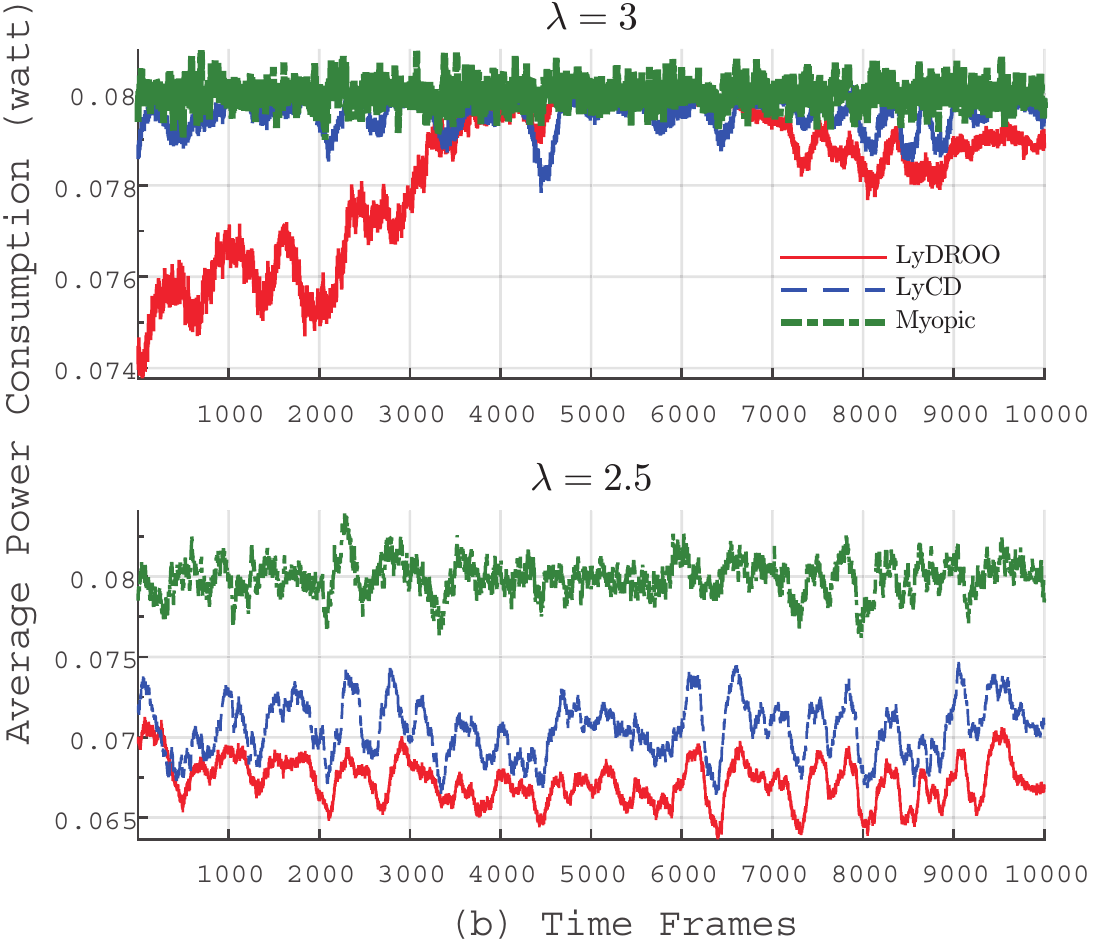}}\quad
  \subfigure{\includegraphics[width=0.23\textwidth]{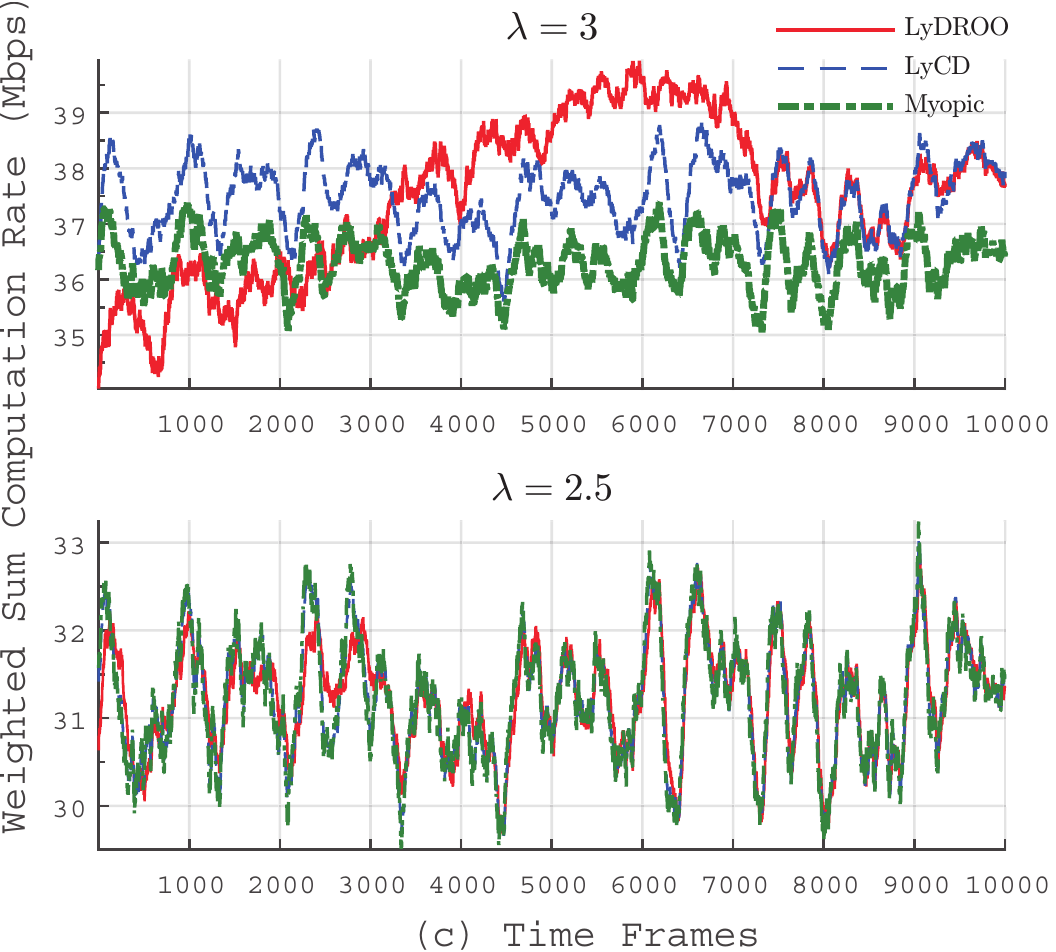}}\quad
  \subfigure{\includegraphics[width=0.25\textwidth]{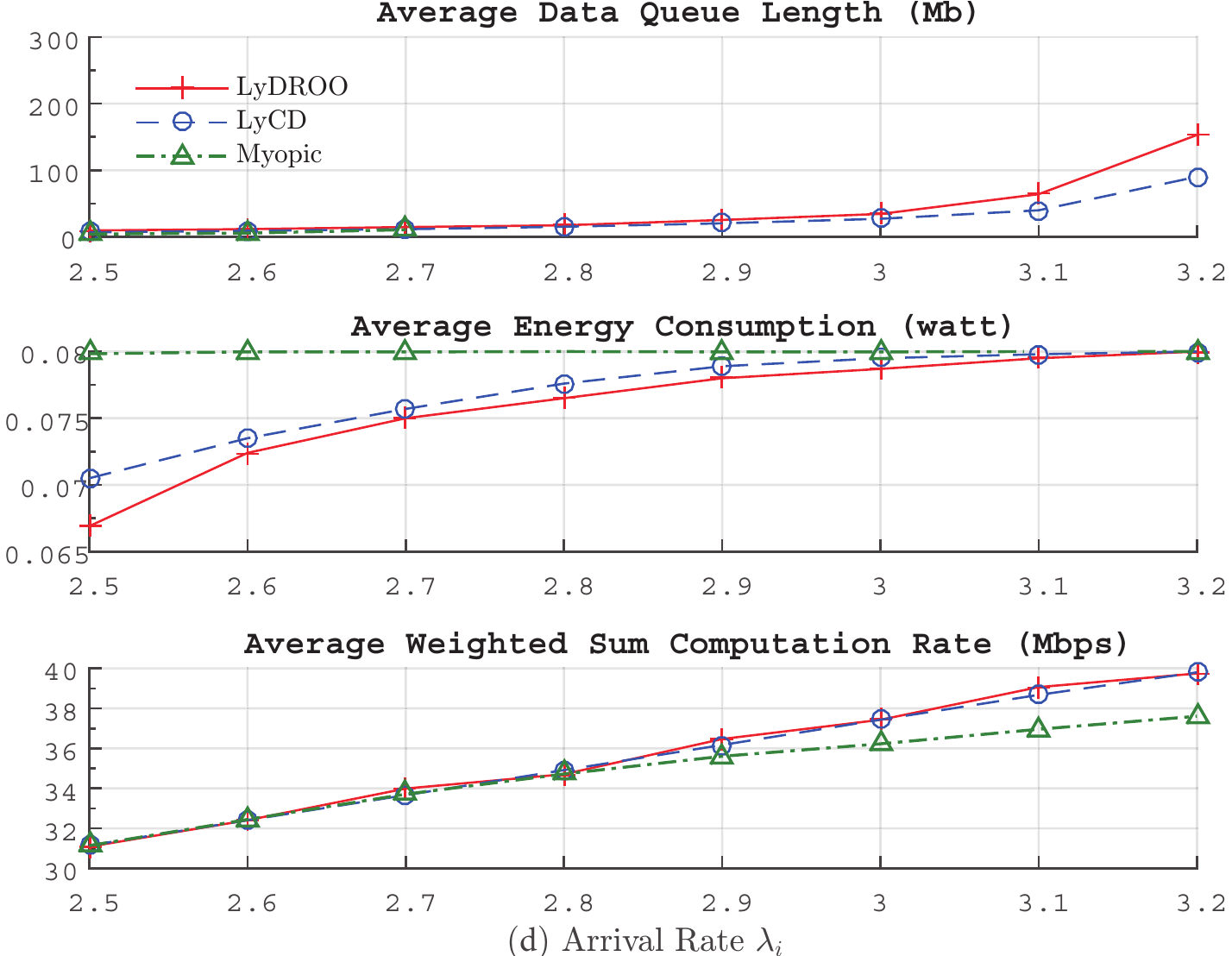}}
  \caption{Convergence performance (a)-(c) and the impact of data arrival rate $\lambda_i$ (d) of different schemes.}
  \label{103}
\end{figure*}

\section{Simulations}
In this section, we use simulations to evaluate the performance of the proposed LyDROO algorithm. All the computations are evaluated on a TensorFlow 2.0 platform with an Intel Core i5-4570 3.2GHz CPU and 12 GB of memory. We assume that the average channel gain $\bar{h}_i$ follows a path-loss model $\bar{h}_i = A_d\left(\frac{3\times 10^8}{4\pi f_c d_i}\right)^{d_e}$, $i=1,\cdots,N$, where $A_d =3$ denotes the antenna gain, $f_c =915$ MHz denotes the carrier frequency, $d_e =3$ denotes the path loss exponent, and $d_i$ in meters denotes the distance between the $i$th WD and the ES. $h_i$ follows an i.i.d. Rician distribution with line-of-sight link gain equal to $0.3 \bar{h}_i$. The noise power $N_0 = W \upsilon_0$ with noise power spectral density $\upsilon_0 = -174$ dBm/Hz. Unless otherwise stated, we consider $N=10$ WDs equally spaced with $d_i = 120 + 15(i-1)$, for $i=1,\cdots,N$. The weight $w_i=1.5$ if $i$ is an odd number and $w_i=1$ otherwise. Other parameter values are set as: $T=1$ second, $W = 2$ MHz, $f_i^{max} = 0.3$ GHz, $P_i^{max} = 0.1$ watt, $\gamma_i = 0.08$ watt, $V=20$, $\phi = 100$, $\nu = 1000$, $q = 1024$, $\delta_T = 10$, $\delta_M=32$, $|\mathcal{S}^t|= 32$. For performance comparison, we consider two benchmark methods: 1) Lyapunov-guided Coordinate Decent (LyCD): The only difference with LyDROO is that LyCD applies the coordinate decent (CD) method \cite{2018:Bi} to solve (\ref{7}), which iteratively applies one-dimensional search to update the binary offloading decision vector $\mathbf{x}^t$. We have verified through extensive simulations that the CD method achieves close-to-optimal performance of solving (\ref{7}). Therefore, we consider LyCD as a target benchmark of LyDROO. 2) Myopic optimization \cite{2019:Huang}: The Myopic method neglects the data queue backlogs and maximizes the weighted sum computation rate in each time frame $t$ by solving
    \begin{equation*}
    \label{97}
    \small
   \begin{aligned}
    & \underset{\mathbf{x}^t, \boldsymbol{\tau}^t,\mathbf{f}^t,\mathbf{e}_O^t, \mathbf{r}_{O}^t}{\text{maximize}} & &   \mathsmaller\sum_{i=1}^N c_i r^{t}_{i} \\
    & \text{subject to} &  & (\ref{25})-(\ref{85}),\  \ e_i^t \leq t\gamma_i - \mathsmaller\sum_{l=1}^{t-1} e_i^l, \ \forall i,
   \end{aligned}
\end{equation*}
where $e_i^t \leq t\gamma_i - \mathsmaller\sum_{l=1}^{t-1} e_i^l$ ensures that the average power constraint in (\ref{63}) is satisfied up to the $t$th time frame and $\left\{e_i^l | l<t\right\}$ is the known past energy consumptions.

In Fig.~\ref{103}(a)-(c), we evaluate the convergence of proposed LyDROO and the two benchmark methods. We consider two data arrival rates with $\lambda_i = 2.5$ and $3$ Mbps for all $i$, and plot the average data queue length, and average power consumption, and weighted sum computation rate over time. We consider i.i.d. random realizations in $10,000$ time frames, where each point in the figure is a moving-window average of $200$ time frames. In Fig.~\ref{103}(a), we observe that for a low data arrival rate $\lambda_i=2.5$, all the schemes maintain the data queues stable and achieve similar computation rate performance. Besides, they all satisfy the average power constraint $0.08$ W in Fig.~\ref{103}(b). Meanwhile, they achieve the identical rate performance in Fig.~\ref{103}(c). When we increase $\lambda_i$ to $3$, the average data queue length of the Myopic method increases almost linearly with time, indicating that the data arrival rate has surpassed the computation capacity (i.e., achievable sum computation rate). On the other hand, both the LyCD and LyDROO methods can stabilize the data queues. In between, the LyCD method maintains lower data queue length over all time frames. The data queue length of LyDROO experiences quick increase when $t\leq 3,000$. However, as the embedded DNN gradually approaches the optimal policy, it quickly drops and eventually converges to the similar queue length and rate performance as the LyCD method after around $t=7,500$, indicating its fast convergence even under highly dynamic queueing systems.

In Fig.~\ref{103}(d), we vary the data arrival rate $\lambda_i$ from $2.5$ to $3.2$ Mbps, and plot the performance after convergence (points with infinite data queue length are not plotted). We omit the results for $\lambda_i>3.2$ because we observe that none of the three schemes can maintain queue stability, i.e., arrival rate surpasses the achievable sum computation rate. All the three schemes satisfy the average power constraints under different $\lambda_i$. The data queues are stable with LyCD and LyDROO under all the considered $\lambda_i$, while the queue lengths of the Myopic scheme become infinite when $\lambda_i > 2.7$. The results show that both LyDROO and LyCD achieve much larger stable capacity region than the Myopic method, and thus are more robust under heavy workload. We also observe that LyCD and LyDROO achieve identical computation rate performance in all the considered cases. This is because when the data queues are long-term stable, the average computation rate of the $i$th WD (departures rate of the data queue) equals the data arrival rate $\lambda_i$, and thus the achievable average weighted sum computation rate is $\sum_{i=1}^N c_i\lambda_i$ for both schemes. In fact, this also indicates that both LyDROO and LyCD achieve the \emph{optimal computation rate} performance in all the considered setups.

We further examine the computational complexity. Here, we consider a fixed total network workload $30$ Mbps and equally allocate $\lambda_i = 30/N$ to each WD for $N\in\left\{10,20,30\right\}$. The locations of the $N$ WDs are evenly spaced within $[120,255]$ meters distance to the ES. The two LyDROO and LyCD method achieve similar computation rate performance for all $N$ and all the long-term constraints are satisfied. In terms of execution delay, LyDROO and LyCD take on average $\left\{0.021,0.108,0.156\right\}$ and $\left\{0.27,2.57,8.02\right\}$ seconds to generate an offloading decision for $N\in\left\{10,20,30\right\}$, respectively. LyCD consumes acceptable latency when $N=10$, but significantly long latency when $N=30$, e.g., around $50$ times longer than that of LyDROO. Because the channel coherence time of a common indoor IoT system is no larger than several seconds, the long execution latency makes LyCD costly even infeasible in a practical MEC system with online offloading decision. The proposed LyDROO algorithm, in contract, incurs very short latency overhead, e.g., around $3\%$ overhead when the time frame is $5$ seconds for $N=30$. Therefore, the LyDROO algorithm can be efficiently applied in an MEC system under fast channel variation.

\section{Conclusions}
In this paper, we have studied an online stable computation offloading problem in a multi-user MEC network under stochastic wireless channel and task data arrivals. We formulated a multi-stage stochastic MINLP problem that maximizes the average weighted sum computation rate of all the WDs under long-term queue stability and average power constraints. To tackle the problem, we proposed a LyDROO framework that combines the advantages of Lyapunov optimization and DRL. We show that the proposed approach can achieve optimal computation rate performance meanwhile satisfying all the long-term constraints. Besides, its incurs very low execution delay in generating an online action, and converges within relatively small number of iterations.

\end{document}